\documentclass[longauth]{aa} 
\usepackage{txfonts}
\usepackage[utf8]{inputenc}
\usepackage[T1]{fontenc}
\usepackage{mathtools}\usepackage{mathtools}
\usepackage{bm}
\usepackage{graphicx}
\usepackage{color}
\usepackage{colortbl}
\usepackage{hyperref}
\hypersetup{colorlinks=true,allcolors=[rgb]{0,0.5,0.5}} 

\makeatletter
\renewcommand*\aa@pageof{, page \thepage{} of \pageref*{LastPage}}

\usepackage{soul}

\definecolor{teal}{rgb}{0.0, 0.5, 0.5}

\newcommand{\sgr}{Sgr~A$^{*}$~}
\newcommand{\Msol}{$M_{\odot}$\,}

\graphicspath{{./}{Figures/}}

\begin{document} 

    \title{Where intermediate-mass black holes could hide \\ in the Galactic Centre}
    \subtitle{A full parameter study with the S2 orbit}
    
    \author{The GRAVITY\thanks{
    GRAVITY is developed in a collaboration by MPE, LESIA of Paris Observatory / CNRS / Sorbonne Université / Univ. Paris Diderot and IPAG of Université Grenoble Alpes / CNRS, MPIA, Univ. of Cologne, CENTRA - Centro de Astrofisica e Gravitação, and ESO \hspace{2cm} * Corresponding author: O. Straub (ostraub@mpe.mpg.de)
    } Collaboration:
          O. Straub\inst{1,2},
          M. Bauböck\inst{3,1},
          R. Abuter\inst{4},
          N. Aimar\inst{5}, 
          P. Amaro Seoane\inst{19,1,20}
          A. Amorim\inst{6,7},
          J.P. Berger\inst{8,4}, 
          H. Bonnet\inst{4}, 
          G. Bourdarot\inst{1},
          W. Brandner\inst{9}, 
          V. Cardoso\inst{7,22},
          Y. Clénet\inst{5}, 
          Y. Dallilar\inst{1},
          R. Davies\inst{1},
          P.T. de Zeeuw, 
          J. Dexter\inst{11}, 
          A. Drescher\inst{1}, 
          F. Eisenhauer\inst{1}, 
          N.M. Förster Schreiber\inst{1}, 
          A. Foschi\inst{7,10},
          P. Garcia\inst{12,7}, 
          F. Gao\inst{13,1}, 
          E. Gendron\inst{5}, 
          R. Genzel\inst{1,14}, 
          S. Gillessen\inst{1}, 
          M. Habibi\inst{1}, 
          X. Haubois\inst{15}, 
          G. Heißel\inst{21,5}, 
          T. Henning\inst{9}, 
          S. Hippler\inst{9}, 
          M. Horrobin\inst{17}, 
          L. Jochum\inst{15}, 
          L. Jocou\inst{8}, 
          A. Kaufer,\inst{15} 
          P. Kervella\inst{5}, 
          S. Lacour\inst{5}, 
          V. Lapeyrère\inst{5}, 
          J.-B. Le Bouquin\inst{8}, 
          P. Léna\inst{5}, 
          D. Lutz\inst{1}, 
          T. Ott\inst{1}, 
          T. Paumard\inst{5}, 
          K. Perraut\inst{8}, 
          G. Perrin\inst{5}, 
          O. Pfuhl\inst{4,1}, 
          S. Rabien\inst{1}, 
          D. C. Ribeiro\inst{1},
          M. Sadun Bordoni\inst{1},
          S. Scheithauer\inst{9}, 
          J. Shangguan\inst{1},
          T. Shimizu\inst{1},
          J. Stadler\inst{16,1}, 
          C. Straubmeier\inst{17}, 
          E. Sturm\inst{1}, 
          L.J. Tacconi\inst{1}, 
          F. Vincent\inst{5}, 
          S. von Fellenberg\inst{18,1}, 
          F. Widmann\inst{1}, 
          E. Wieprecht\inst{1}, 
          E. Wiezorrek\inst{1}, 
          J. Woillez\inst{4}
          }

    \institute{
        Max Planck Institute for Extraterrestrial Physics, Giessenbachstraße 1, 85748 Garching, Germany
        \and
        ORIGINS Excellence Cluster, Boltzmannstraße 2, D-85748 Garching, Germany
        \and
        Department of Physics, University of Illinois, 1110 West Green Street, Urbana, IL 61801, USA
        \and
        European Southern Observatory, Karl-Schwarzschild-Straße 2, 85748 Garching, Germany
        \and 
        LESIA, Observatoire de Paris, Université PSL, CNRS, Sorbonne Université, Université de Paris, 5 place Jules Janssen, 92195 Meudon, France
        \and 
        Universidade de Lisboa – Faculdade de Ciências, Campo Grande, 1749-016 Lisboa, Portugal
        \and 
        CENTRA - Centro de Astrofísica e Gravitação, IST, Universidade de Lisboa, 1049-001 Lisboa, Portugal
        \and 
        Univ. Grenoble Alpes, CNRS, IPAG, 38000 Grenoble, France 
        \and
        Max Planck Institute for Astronomy, Königstuhl 17, 69117 Heidelberg, Germany 
        \and
        Universidade do Porto, Faculdade de Engenharia, Rua Dr. Roberto, Frias, 4200-465 Porto, Portugal
        \and 
        Department of Astrophysical \& Planetary Sciences, JILA, Duane Physics Bldg., 2000 Colorado Ave, University of Colorado, Boulder, CO 80309, USA
        \and 
        Faculdade de Engenharia, Universidade do Porto, rua Dr. Roberto Frias, 4200-465 Porto, Portugal
        \and 
        Hamburger Sternwarte, Universität Hamburg, Gojenbergsweg 112, 21029 Hamburg, Germany
        \and 
        Departments of Physics and Astronomy, Le Conte Hall, University of California, Berkeley, CA 94720, USA
        \and 
        European Southern Observatory, Casilla 19001, Santiago 19, Chile
        \and 
        Max Planck Institute for Astrophysics, Karl-Schwarzschild-Straße 1, D-85748 Garching, Germany
        \and 
        $1^{st}$ Institute of Physics, University of Cologne, Zülpicher Straße 77, 50937 Cologne, Germany
        \and
        Max Planck Institute for Radio Astronomy, auf dem H\"ugel 69, D-53121 Bonn, Germany
        \and
        Institute of Multidisciplinary Mathematics, UPV, València, Spain
        \and
        Kavli Institute for Astronomy and Astrophysics, Beijing, China
        \and
        Advanced Concepts Team, European Space Agency, TEC-SF, ESTEC, Keplerlaan 1, 2201 AZ Noordwijk, The Netherlands
        \and
        Niels Bohr International Academy, Niels Bohr Institute, Blegdamsvej 17, 2100 Copenhagen, Denmark
         }

    \date{\today}
    
    \abstract{
    %
    In the Milky Way the central massive black hole, \sgr, coexists with a compact nuclear star cluster that contains a sub-parsec concentration of fast-moving young stars called S-stars. Their location and age are not easily explained by current star formation models, and in several scenarios the presence of an intermediate-mass black hole (IMBH) has been invoked.
    }
    {
    We use GRAVITY astrometric and SINFONI, KECK, and GNIRS spectroscopic data of S2, the best known S-star, to investigate whether a second massive object could be present deep in the Galactic Centre (GC) in the form of an IMBH binary companion to \sgr. 
    }
    {
    To solve the three-body problem, we used a post-Newtonian framework and consider two types of settings: (i) a hierarchical set-up where the star S2 orbits the \sgr--~IMBH binary and (ii) a non-hierarchical set-up where the IMBH trajectory lies outside the S2 orbit. In both cases we explore the full 20-dimensional parameter space by employing a Bayesian dynamic nested sampling method. 
    }
    {
    For the hierarchical case we find the strongest constraints: IMBH masses > 2000 \Msol on orbits with smaller semi-major axes than S2 are largely excluded. For the non-hierarchical case, the chaotic nature of the problem becomes significant: the parameter space contains several pockets of valid IMBH solutions. However, a closer analysis of their impact on the resident stars reveals that IMBHs on semi-major axes larger than S2 tend to disrupt the S-star cluster in less than a million years. This makes the existence of an IMBH among the S-stars highly unlikely.
    }
    {
    The current S2 data do not formally require the presence of an IMBH. If an IMBH hides in the GC, it has to be either a low-mass IMBH inside the S2 orbit that moves on a short and significantly inclined trajectory or an IMBH with a semi-major axis $> 1''$. We provide the parameter maps of valid IMBH solutions in the GC and discuss the general structure of our results and how future observations can help to put even stronger constraints on the properties of IMBHs in the GC.
    }
    
 
    \titlerunning{Where IMBHs hide in the GC}
    \authorrunning{The GRAVITY Collaboration}
    \maketitle

\section{Introduction}
\label{sec:intro}
The nuclear star cluster in the Milky Way can, due to its proximity to Earth \citep[$R_0 = 8.28$ kpc, ][]{GRAVITYcol19b, GRAVITYcol21, do+19}, be resolved into individual stars. In its entirety, it has an oblate shape and extends in the K-band to about $178''$ \citep[i.e. 7.2 pc at $R_0$, ][]{bec+68, sch+14, fri+16} around the central massive black hole, Sagittarius $A^*$ \citep[\sgr, ][]{eck+96, ghe+98, sch+02, ghe+08, GRAVITYcol18a} and consists predominantly of old and evolved stars. However, in its innermost region, the central $12''$ (0.5 pc), it contains a dense and diverse population of stars with a surprising accumulation of young and massive O and B stars. They are found in the stellar disc of WR/O stars that extends from $0.8'' - 12''$ and shows a clockwise motion \citep[][]{pau+06, bar+09, lu+09, bar+10, yel+14}, and in the S-star cluster that resides inside the disc's truncation radius and can have ages as young as 3 - 15 $\times \, 10^6$~years \citep[][]{ghe+03, eis+05, pfu+11, lu+13, hab+17, vfe+22}.

Accompanying the morphology of the Galactic core region are two puzzling observations. On the one hand, there are the isotropically oriented orbital planes and the approximately thermal distribution of the orbital eccentricities of the S-stars. With only a few million years of age, the early B-type stars thus appear too young to be that thermally relaxed in such close proximity to \sgr \citep[the paradox of youth;][]{ghe+03}. On the other hand, in dynamically relaxed systems, one would expect mass segregation where more massive bodies like the WR/O stars are located closer to the centre than the less massive S-stars \citep[][]{ale+09}.

Over the past decades, many models have been proposed to explain the age and location of the S-stars. \citet{hm03} were the first to suggest that an intermediate-mass black hole (IMBH) is present in the Galactic Centre (GC). They argue that an IMBH could have dragged the S-stars from a greater, more star formation friendly distance inwards. However, the telltale trail of young stars outside 0.5 pc, which would support a collective inward migration of such a cluster, is not observed \citep{fel+15}. Nonetheless, the idea that an IMBH is associated with the location and the distribution of orbital elements of the S-star has been picked up in a variety of studies, and is still a matter of debate today. 

Not all scenarios require an IMBH, though. \citet{che+14, che+15} resolve the paradox of youth and mass segregation problem with a rapid redistribution of stellar orbits based on a Kozai-Lidov-like resonance induced by a stellar disc that was more massive and extended in the past. \citet{gen+20} argue that if the S-stars are sourced by the WR/O disc via the Hills mechanism \citep[stellar binary disruption by a massive third body;][]{hil88}, an additional relaxation mechanism is needed to reproduce their present-day distribution on the short timescale given by their ages. They conclude that within a few million years either scalar resonant relaxation from the observed isotropic star cluster or an IMBH of $\sim 10^3$~\Msol at 250~mas could achieve the observed eccentricities. Employing a cluster of stellar black holes (SBHs) as relaxation agent, \citet{per+09} found in N-body simulations running over 20~Myrs that a thermal eccentricity distribution is a natural consequence of random gravitational encounters of stars with a population of SBHs with a total mass of $\propto 10^4$~\Msol in the inner 0.1~pc. Assuming a cluster of more massive SBHs, \citet{tep+21} arrive at the same conclusion, but on a shorter timescale. This is consistent with the upper limit on the dark mass distribution of about 15000~\Msol within 0.1-0.2~pc derived by \citet{GRAVITYcol22}. 

The paper is structured as follows. Sections \ref{sec:realitycheck} and \ref{sec:constraints} discuss whether or not it is realistic to expect an IMBH in the GC and what constraints on its mass and location have been found in previous studies. In Section \ref{sec:data} we describe the data set used, and in Section \ref{sec:method} we present the model and methodology we used to fit them. Our results follow in Section \ref{sec:results}. In Section \ref{sec:analysis} we discuss the stability analysis. Finally, in Section \ref{sec:discussion} we add concluding remarks and an outlook on the future.

\section{Possibility of IMBHs in the Galactic Centre}
\label{sec:realitycheck}
Theoretically, black holes can have any mass upwards of the Planck mass.\footnote{$m_P = (\frac{\hbar c}{G})^{1/2} = 2.2\times10^{-5}$~g} Astrophysical black holes, however, essentially only come in two `flavours'. 

The first is 
stellar black holes, with masses ranging from about $3 - 100$~\Msol, where \Msol = $2\times10^{33}$~g, which form via gravitational collapse of massive stars that depleted their nuclear energy source \citep[e.g.][]{opp+39, pen65, mir17}. 

The second flavour is massive black holes (MBHs), with masses higher than $10^6$~\Msol, which are thought to form via direct or indirect gravitational collapse of an initial massive gas cloud and to co-evolve symbiotically with their host galaxies \citep[e.g.][]{ree78}. Although there is an emerging consensus regarding the growth of supermassive BHs thanks to So{\l}tan's argument \citep[][]{sol82}, the evolution of MBHs with masses up to $10^7$~\Msol, such as our own MBH in the Galactic Centre (with a mass of $\sim 4.2 \times 10^6$~\Msol), is enigmatic.

There is compelling evidence for the existence of SBHs from both electromagnetic observations \citep[e.g.][]{nar+13, cas+14, cor+16} and gravitational wave detection \citep{LIGOcol20c}. Equally well established is the occurrence of MBHs at the centres of massive galaxies \citep[e.g.][]{mag+98, vol10, kor+13}. Moreover, the increasing number of observations of luminous quasars at very high redshift indicates that some supermassive BHs with masses $>10^8$~\Msol already existed when the Universe was less than a billion years old \citep{mor+11, wu+15, ban+18, yan+20}.

Intermediate-mass black holes are thought to bridge the gap between these two BH populations and, more importantly, to be the building blocks in the formation process of MBHs. Understanding them is crucial to answering the question of how the young and supermassive quasars could develop into behemoths on such short timescales.\\

The following three MBH formation channels predict the appearance of IMBHs at different times and in different numbers. There are two early formation mechanisms that rely on the properties of zero-metallicity gas and can therefore only operate at redshift z~>~10. In the young Universe, the pristine hydrogen gas could have either coagulated into the first generation of massive Population III stars \citep{mad+01} or it could have contracted uniformly to directly form a single supermassive star that then collapsed into an intermediate-mass seed BH \citep{loe+94, beg+06}, possibly via an accreting quasi-star phase \citep{hoy+63, beg10, wis+19}. The inefficient cooling due to the presence of primordial hydrogen inhibits premature fragmentation and pair-instability supernovae such that the Population III stars and the supermassive star could have reached masses significantly higher than 100~\Msol and lead to early intermediate-mass seed BHs \citep{ohk+09}. 

Quite distinct from the two early seeding mechanisms is the third dynamical formation channel where gravitational runaway and hierarchical black hole mergers in dense nuclear star clusters can form many IMBH kernels \citep{qui+90, por+02, fre+06, sto+17}. \citet{ant+19} have calculated that IMBHs can indeed form via hierarchical mergers in star clusters with high enough escape velocities and densities.  \citet{riz+21} have pointed out that  IMBHs could form in $\lesssim$ 15 Myr, in particular in young
and compact star clusters. While the two early seeding mechanisms produce at most one IMBH per galaxy halo at high redshift, this latter process can operate throughout cosmic time and could provide a channel to create an IMBH in any dense stellar system \citep[for comprehensive reviews, see][]{mil+04, mez17, gre+20}.  Recently, a mass-gap SBH (or low-mass IMBH) of around 150~\Msol has been identified as the product of a coalescence of two SBHs via gravitational wave detection \citep[GW190521, ][]{LIGOcol20a, LIGOcol20b, nit+21}, supporting scenarios with dynamical hierarchical mergers.\\


Today, intermediate-mass black holes that formed via the early seeding processes are thus expected to populate the centres of low-mass dwarf\footnote{Some dwarf galaxies can have surprisingly massive central BHs \citep[e.g.][]{bus+21}, possibly due to dynamical mergers of IMBHs in complexes of young stellar clusters \citep[][]{ama+14}.} and satellite galaxies \citep[e.g.][]{mez+16,mez+18}, whereas IMBHs formed via dynamical mergers are thought to be found rather in globular clusters \citep{mil+02, bau+05} and nuclear star clusters \citep{mil+09, neu+20}. The most convincing IMBH candidates are indeed found in low-mass galaxies and have masses $10^4 \lesssim M < 10^6$~\Msol, for example HLX-1 \citep[][]{far+09, web+17} and the Large Magellanic Cloud \citep[LMC, ][]{erk+19}. During their evolution, galaxies may accrete nearby satellite or dwarf galaxies, which could deposit a substantial number of wandering IMBHs, each surrounded by a stellar system, in the galactic halos \citep[e.g. the Milky Way, ][]{ras+14}. Moreover, centres of galaxies have in principle deep enough potential wells to retain SBH merger products in their nuclear star clusters \citep[see][]{hai+18, fra+21, ros+22}. Therefore, it seems conceivable that the centre of the Milky Way could host an IMBH. Although the question arises of whether it could hide among the S-stars.

\section{Constraints on IMBH mass and location in the Galactic Centre}
\label{sec:constraints}
The first constraints on the mass and location of an IMBH in the GC came from a study of dynamical processes that can eject hyper-velocity stars from the GC at average speeds of 400-2000 km s$^{-1}$ \citep{yt03} and the measurement of the proper motion of \sgr that is consistent with no acceleration \citep{rb04, rb20}. These studies exclude in essence IMBH masses of $M \gtrsim 3\times10^4$~\Msol within the S-star cluster and the WR/O disc. 

\citet{mer+09} employed long-term N-body simulations to show that the presence of an IMBH can randomise the orbital planes of 19 S-stars in one million years if the IMBH mass exceeds 1500~\Msol and its pericentre distance is smaller than 250~mas. N-body simulations of the orbits of S-stars around \sgr in the presence of an IMBH have been used to study the effects of an IMBH on the orbit of S2 in particular. These codes typically solve the N-body problem numerically (e.g. with a post-Newtonian approximation\footnote{The post-Newtonian approximation is a method for solving Einstein’s field equations by expanding them in terms of a small parameter $\epsilon \sim (v/c)^2$.}) up to order 2.5 and with 21 S-stars in addition to the MBH and the IMBH \citep[][]{gm09, ggm10}. 

Many-body systems are chaotic in nature, and in order to make the orbital fitting procedure manageable the N-body codes traditionally rely on a discrete but serviceable set of reasonable IMBH orbital parameters, for instance three different eccentricity values paired with a range of interesting IMBH masses and a fixed set of inclinations and orbital angles. Another way to tackle the chaotic nature of the three-body problem is used by \citet{nao+20} who studied a high-order analytic approximation of the inverse Kozai-Lidov equations. Considering the stability of the S2 orbit, they could rule out a $10^5$ \Msol companion on a circular orbit with a   semi-major axis greater than 20~mas.

In \citet{GRAVITYcol20} we collected the available constraints on the IMBH mass and semi-major axis in the literature and presented them together with an estimate of the constraints that could be achieved by the GRAVITY instrument. In this work we show the actual IMBH constraints based on GRAVITY (and SINFONI/KECK/GNIRS) data of S2. In terms of simulation and fitting technique, in this work we go a step further than previous N-body simulations and explore not only a few selected sets of IMBH orbits, but the full-dimensional parameter space. In this way we obtain the most realistic constraints based on current high angular resolution interferometric and spectroscopic infrared observations.

\section{Data}
\label{sec:data}
The star S2 moves on a highly elliptical 16-year orbit around \sgr and has been monitored since 1992. The resulting high-precision data of nearly 2.5 orbits have not only lead to the direct measurement of the compact mass in the GC, $M_0 \approx 4.30 \times 10^6$~\Msol, and its distance, $R_0 \approx 8.28$~kpc \citep[][]{GRAVITYcol19b, GRAVITYcol22}, but have also delivered evidence for relativistic effects such as the gravitational redshift \citep[][]{GRAVITYcol18a, do+19} and the Schwarzschild precession  \citep[][]{GRAVITYcol20}, as well as the local position invariance \citep[][]{GRAVITYcol19a}.

In this work, we use the astrometry data taken from 2017--2021 by the GRAVITY beam combiner, a K-band infrared interferometer at the European Southern Observatory's Very Large Telescope (ESO's VLT) together with spectroscopic data collected from 2000--2021 by NIRC2 at the Keck Observatory, SINFONI at the VLT, and GNIRS at the Gemini Observatory \citep[see][for a more detailed description]{GRAVITYcol22}. 

All GRAVITY data have been recorded in low resolution and split (linear) polarisation. Each exposure consists of a total integration time of 320 seconds, comprised of 32 consecutive frames every 10 seconds. One VLT observation block contains two different targets, the star S2 and the black hole \sgr. During the pericentre passage of S2 from 2017 to 2018, both objects were detected simultaneously in the same fibre field of view (FoV = 60 mas). In all epochs from 2019 onwards, the separation between S2 and \sgr has been larger than the FoV and the objects have been targeted individually. In this  dual-beam mode we first take an exposure with the fibre centred on S2 and then dither to \sgr and take a sequence of four exposures. We repeat this 1+4 pattern throughout the available night. We then use the latest version of the standard GRAVITY data reduction pipeline to reduce all data. The interferometric observables, the closure phase and visibility, of the star S2 are consistent with a single point source such that we can use it as a phase reference to calibrate the \sgr exposures. In this way we can calculate the separation vector between S2 and \sgr from the fitted phase offsets \citep[see Appendix A in ][]{GRAVITYcol22}. The resulting GRAVITY astrometry has a root mean square (rms) uncertainty of $\approx 50~\mu$as; SINFONI and KECK -- GNIRS data have a rms uncertainty of  $\approx$ 12 km/s and $\approx$ 45~km/s, respectively.

In this work, we are not using any adaptive optics (AO) astrometric data collected by NACO/VLT. The reason we omit about 75 a priori perfectly valid AO imaging measurements between 2003 and 2019 is that the calibration of the reference frame between NACO and GRAVITY is largely degenerate with adding an IMBH. In sampling such a posterior, the solutions run away towards an arbitrary calibration factor and arbitrarily high IMBH masses. We avoid the problem by excluding the AO measurements and using only the GRAVITY high-resolution interferometric astrometry, which is internally self-consistent and of a much higher precision than the NACO data (rms of about 1.7 mas).

\section{Methodology}
\label{sec:method}
We consider two scenarios. In the hierarchical set-up, \sgr has a close IMBH binary companion with a small semi-major axis $0.01'' \leq a_{\rm i} \leq 0.1''$. The star S2 with $a = 0.125''$ orbits around this   binary's centre of mass. The IMBH orbit lies in this case inside the S2 orbit. In the non-hierarchical set-up the IMBH has a semi-major axis $0.1'' \leq a_{\rm i} \leq 1''$, which crosses the S2 orbit or lies entirely outside of the S2 orbit but still within the S-star cluster. In this set-up the centre of mass is \sgr. We treat these two distinct cases separately.

\subsection{Orbital integration}
\label{sec:integration}
To simulate the orbits of a three-body system consisting of \sgr, the star S2, and an IMBH, we adapted the publicly available REBOUND N-body code \citep{rebound}. We used REBOUND in combination with the REBOUNDx package \citep{reboundx} which incorporates the first-order post-Newtonian effects from all massive bodies in the system. The simulations were integrated using a $15{\rm th}$ order Gauss-Radau integrator \citep[IAS15; ][]{reboundias15}. 

We first add \sgr at the origin of the coordinate system. In order to minimise the error introduced to the S2 orbital parameters due to the transformation between a flat Cartesian coordinate system and the relativistic spacetime around the black hole, we add the star S2 near the apocentre of its orbit (i.e. we set the initial timestamp of the osculating Keplerian orbit to $t_0 = 2010.0$). We then integrate the orbit forward to the date of the last GRAVITY observation used in this work: $t = 2021.570283$. Here we convert the orbital elements of S2 ($a_{\rm S2}$, $e_{\rm S2}$, $i_{\rm S2}$, $\Omega_{\rm S2}$, $\omega_{\rm S2}$, $T_{\rm peri, S2}$) into a state vector consisting of the position and velocity. This ensures the correct starting position with regard to the observational data. We then remove the star S2 and add the IMBH, and redefine the coordinate system so that the origin is now at the centre of mass. Finally, we add the star S2 with the starting position and velocity vectors calculated previously.

Once we have initialised the simulation, we integrate the orbits of all three masses backwards in time to the earliest velocity data point, at $t = 2000.476$. Given the larger uncertainties of the early data points, we integrate backwards in time to make sure S2 is on the correct orbit in the present day. We take into account the Rømer delay arising from the change in the light travel time at various points along the S2 orbit. We approximate the delay following \cite{GRAVITYcol18a} as
\begin{equation}
    t_{\rm em} = t_{\rm obs} - \frac{z(t_{\rm obs})}{c} \left( 1 - \frac{v_z(t_{\rm obs})}{c}\right)
    \label{eq:roemer}
,\end{equation}
where $t_{\rm em}$ is the time at which a photon is emitted, $t_{\rm obs}$ is the time at which it is observed, and $z$ and $v_z$ represent the line-of-sight distance and velocity, respectively. For each observation we therefore first calculate the position and velocity at the observed time and then use these values to approximate the emitted time. We then integrate the orbit of S2 to $t = t_{\rm em}$ to compare with the data. 

The REBOUNDx module includes general relativistic effects up to first order in the post-Newtonian approximation in the calculation of the orbits of all three masses, but it does not account for the relativistic effects experienced by the photons emitted by those masses. We therefore explicitly account for the transverse Doppler shift and the gravitational redshift for the star S2 when calculating the observed radial velocity. Specifically, we assume a Schwarzschild geometry for the MBH \sgr and an observer at infinity. This allows us  to calculate the approximated observed radial velocity by multiplying the two correction terms, respectively, which leads to 
\begin{equation}
    v_{\rm obs} = v_z + \left(1 - \gamma \right) + \left(1 - \sqrt{1 - \frac{r_S}{r}} \right), 
\end{equation}
where $\gamma$ is the Lorentz factor and $r_S$ is the Schwarzschild radius.

We can then calculate a $\chi^2$ value by comparing the model orbital motion of S2 to the observed data. For the spectral velocity measurement, the measured quantity is simply $v_{\rm obs}$ calculated above. For the astrometric position, the relevant quantities to compare to the GRAVITY measured separation between S2 and the emission from \sgr are the modelled differences in right ascension and declination $\left(RA_{\rm S2} - RA_{\rm SgrA^*} \, , \,  DEC_{\rm S2} - DEC_{\rm SgrA^*}\right)$.

\subsection{Posterior sampling}
\label{sec:posteriors}
Once we are able to calculate a $\chi^2$ value for any point in the parameter space, we turn to sampling methods to evaluate the posterior. Since the general three-body problem is chaotic, the orbit of S2 can depend very sensitively on the IMBH orbital parameters. If the two masses interact significantly, S2 will deviate widely from the observed orbit. This leads to a complex posterior distribution that features many local maxima and degeneracies between parameters. 

We use dynamic nested sampling \citep{ski04, ski06, hig19} as implemented by the \texttt{dynesty} code \citep{dy_main} to calculate both the posterior distribution and the model evidence. Dynamic nested sampling is a  generalisation of the nested sampling algorithm, which dynamically adjusts the number of samples taken in different regions of the parameter space in order to maximise calculation accuracy. We use this approach for two principal reasons. First, nested sampling is better able to capture multi-modal posterior distributions than more traditional Markov chain Monte Carlo methods \citep[see e.g.][]{ash+22}. Second, nested sampling directly calculates the evidence, allowing for model comparison (in this case between scenarios with and without an IMBH) as well as parameter constraints.

After some experimentation, we have found that a nested sampling run with at least 8000 live points is needed to reproducibly converge on the posterior distribution. We have found the best numerical performance using the `rwalk' sampling method and the `multi' bounding distribution \citep[see][for details]{fer09, ski06}. In order to ensure that we have explored the full parameter space, we explored independent runs with different initialisation parameters or negligibly different boundaries as well as a run with 16000 initial live points. We find that all produce a nearly identical posterior distribution. 

To confirm the accuracy of our set-up, we compare the posterior distributions of the S2 orbital parameters as well as the mass and distance of \sgr with the published values. We recover the published values to within the error bars in both a fiducial run without an IMBH as well as a full run with free IMBH orbital parameters. We also recover the expected degeneracies between  the mass and distance of \sgr.

\begin{table}[ht]
    \centering
    \begin{tabular}{| l | c | l |}
     \hline
     Parameter     &   Starting Point   & Boundaries \\
     \hline\hline
     $M_0$ (\Msol)        & $4.2\times10^6$ & $\pm 5\times10^5$ \\
     \hline
     $R_0$ (kpc)          & 8.25            & $\pm$ 1.0 \\
     \hline
     $v_{\rm z, 0}$ (km/s)         & 0               & $\pm$ 5  \\
     \hline
     $a_{\rm S2}$ ($''$)            & 0.125           & $\pm$ 0.02 \\
     \hline
     $e_{\rm S2}$                   & 0.87            & $\pm$ 0.05\\
     \hline
     $i_{\rm S2}$ ($^\circ$)        & 134             & $\pm$ 5 \\
     \hline
     $\Omega_{\rm S2}$ ($^\circ$)   &  228            & $\pm$ 5 \\
     \hline
     $\omega_{\rm S2}$ ($^\circ$)   &  66             & $\pm$ 5  \\
     \hline
     $T_{\rm peri, S2}$ (y)         & 2018.4          & $\pm$ 0.2 \\
     \hline
     $M_{\rm i}$ (\Msol)            & 5010            & $\pm$ 5000 \\
     \hline
     $a_{\rm i}$ ($''$)             & 0.51            & $\pm$ 0.5 \\
     \hline
     $e_{\rm i}$                    & 0.48            & $\pm$ 0.47\\
     \hline
     $i_{\rm i}$ ($^\circ$)         & 180             & $\pm$ 180 \\
     \hline
     $\Omega_{\rm i}$ ($^\circ$)    & 180             & $\pm$ 180\\
     \hline
     $\omega_{\rm i}$ ($^\circ$)    & 180             & $\pm$ 180  \\
     \hline
     $\mu_{\rm i}$ ($^\circ$)       & 180             & $\pm$ 180  \\
     \hline
    \end{tabular}
    \vspace{0.25cm}
    \caption{Fitting parameters, their initial values, and the boundaries. Not listed are the \sgr parameters ($x_0, y_0, z_0, v_{\rm x, 0}, v_{\rm y, 0}$), which are also allowed to vary.}
    \label{tab:free_param}
\end{table}

The parameters of our simulation are summarised in Table~\ref{tab:free_param}. Along with the mass and the six orbital parameters of the IMBH, we allow the orbital parameters of S2, the mass and distance of \sgr, and a global velocity offset to vary. We   chose to parametrise the initial position of the S2 orbit with the time of pericentre passage $T_{\rm peri}$, which is well constrained from observations. In order to limit the duplication of IMBH orbits, however, we use the mean anomaly at $t_0$ to parametrise its initial position, which naturally confines the initial conditions to a single orbital period. 

\begin{figure*}[t]
    \centering
    \includegraphics[width=0.9\columnwidth]{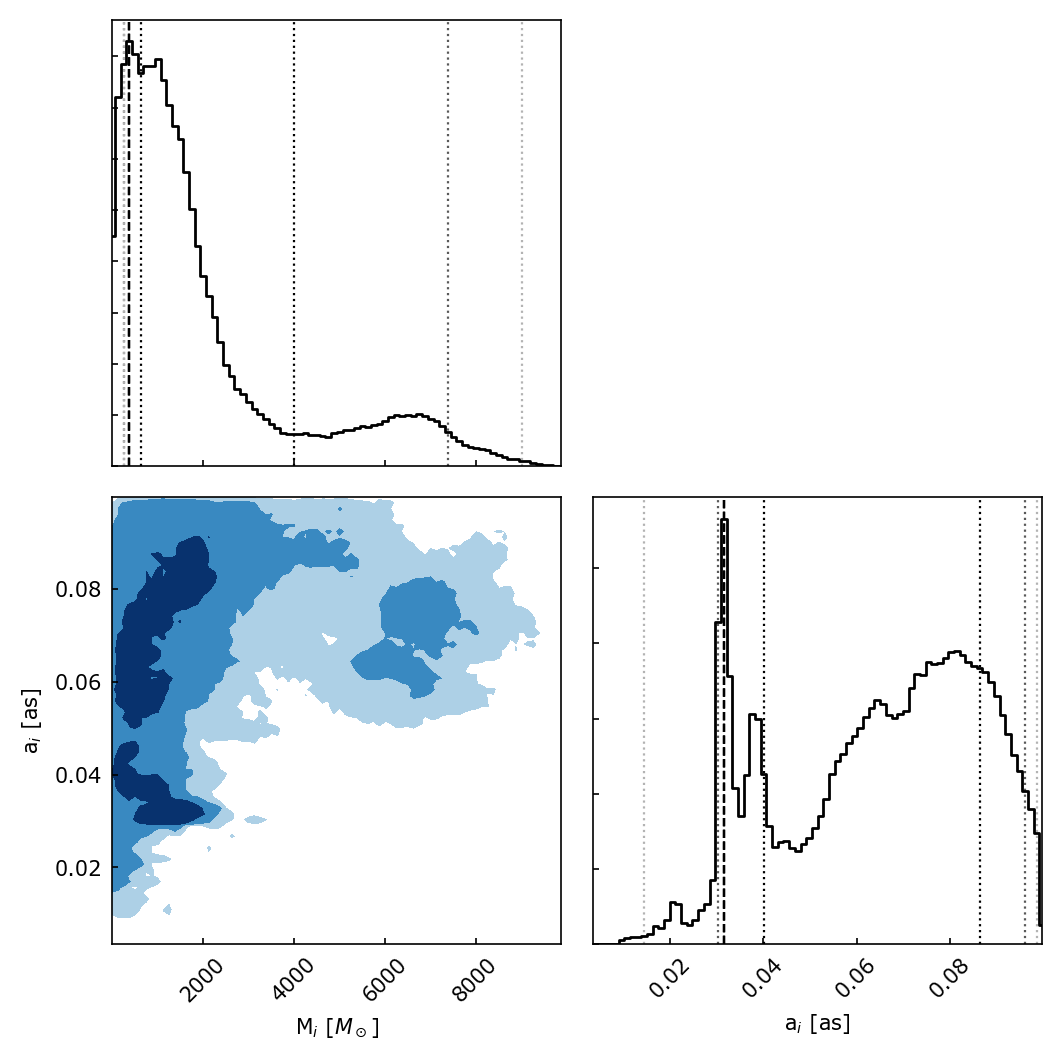}
    \hfill
    \includegraphics[width=0.9\columnwidth]{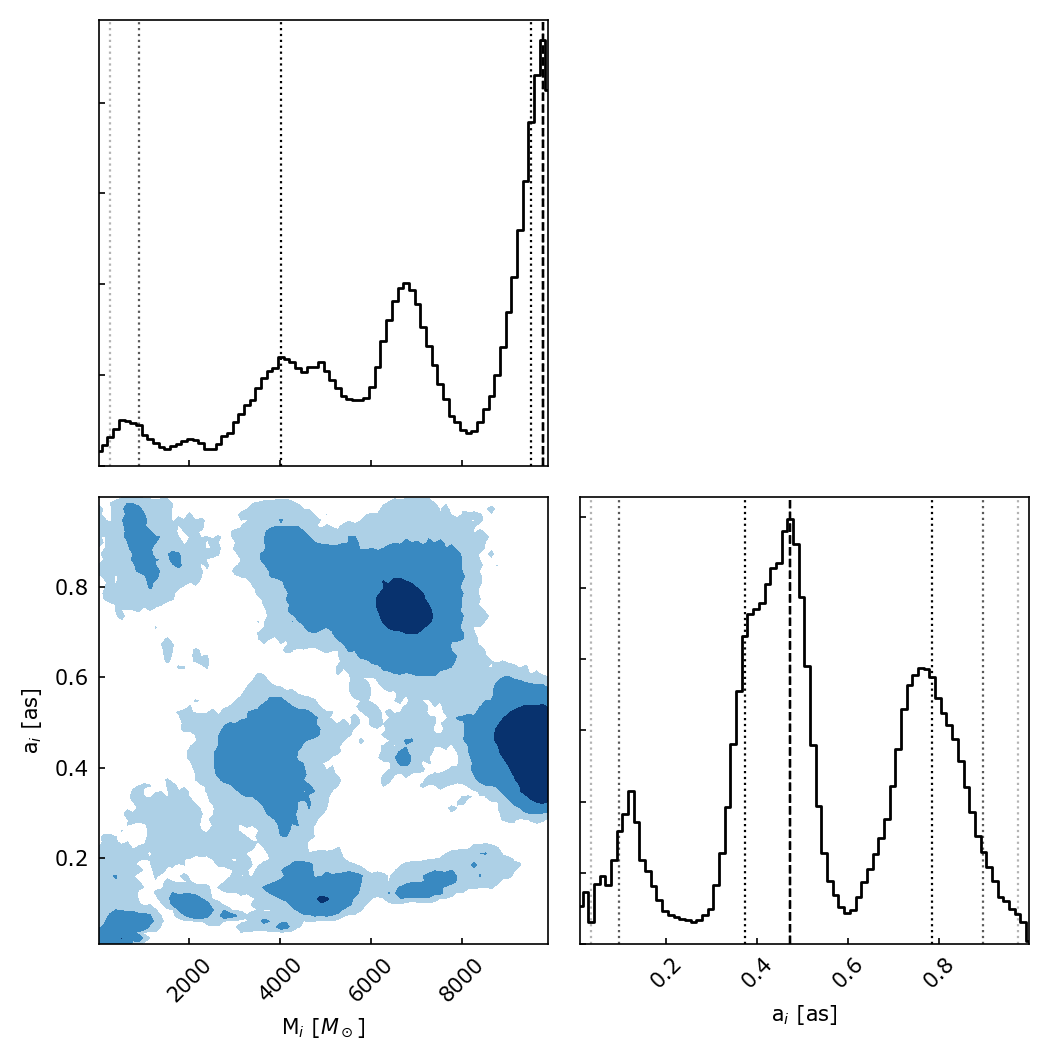}
    \caption{Posterior distributions of IMBH orbits. Left: Posterior distribution over the mass and semi-major axis of the IMBH for orbits with a semi-major axis smaller than 0.1". The contours correspond to 39\%, 86\%, and 98.8\% (from dark blue to light blue) enclosed likelihood. Right: Same as left, but  for orbits extending out to a semi-major axis of $1''$.}
    \label{fig:constraints_small_large}
\end{figure*}

Table~\ref{tab:free_param} also shows the initial value and allowed range for each parameter. We use a flat prior across the space (-range, +range). The S2 orbital parameters and \sgr mass and distance are already tightly constrained by previous fits to (partly) the same data as used here. We adopt values close to \citet{GRAVITYcol18a}, with a range scaled from the errors quoted therein. For the IMBH we allow the angular orbital parameters to vary between 0$^\circ$ and 360$^\circ$. We expect to have the greatest discriminating power for IMBHs that lie within or close to the S2 orbit, as the potential for three-body interactions is thus maximised. However, the minimum time step to accurately calculate orbits decreases as the closest approach distance decreases. This decreased time step increases the computational time for each likelihood evaluation. Given that our results depend on a robust exploration of the parameter space, we therefore choose an initial range of semi-major axes between $0.01''$ and $0.1''$ and limit eccentricities to be less than $0.95$. With this set-up, a complete run of the parameter estimation can be completed on a moderately sized cluster (60 cores) within approximately one week. We additionally explore the non-hierarchical scenario in a second run where we allow the IMBH semi-major axis to extend out to $1''$.

\section{Constraints on IMBHs in the GC from the S2 orbit}
\label{sec:results}
From the posterior sampling we obtain the full set of IMBH orbital parameters (see Appendix~\ref{app:fullparam}). The left panel of Fig.~\ref{fig:constraints_small_large} shows the posterior distribution of the IMBH mass and the semi-major axis of its orbit for a prior range of $a_{\rm i} < 0.1''$. For all IMBH semi-major axes inside the S2 orbit, we exclude IMBH masses greater than $4010 M_\odot$ at the 86\% level. At small semi-major axes $\lesssim 0.05''$, these limits are considerably stronger, and IMBHs with a mass greater than $\approx 2000 M_\odot$ are very strongly excluded.

We find a global minimum $\chi^2$ value of 219.53 for an IMBH with a mass of 1904~\Msol and a semi-major axis of $0.031''$, compared to a minimum $\chi^2$ of 224.1 for an S2-only model. 
Since the IMBH model formally fits the data better than the S2-only model, we calculate the evidence for each model by integrating over the posterior distribution. We find that the log-evidence for the two models are essentially identical: $\log(z) = 124.80$ for S2-only, and $\log(z) = 124.79$ for the IMBH. We therefore conclude that we cannot distinguish between these models and that our constraints quoted above are indeed upper limits.

The right panel of Fig.~\ref{fig:constraints_small_large} shows the posterior distribution of the IMBH mass and the semi-major axis of its orbit for a prior range of $0.1'' < a_{\rm i} < 1''$. Here we find a minimum $\chi^2$ value of 220.54 for an IMBH with a mass of 5842~\Msol and a semi-major axis of $0.164''$. However, the posterior peaks at the upper edge of the prior mass range, implying that we do not generate a valid upper limit on the mass or a constraint on the semi-major axis. These peaks in the posterior correspond to an IMBH on a large orbit that essentially does not interact with S2 over the $\approx$20-year timescale probed here, rendering it undetectable with our current method.

We find the shape of the allowed region in the $M_{\rm i} - a_{\rm i}$ parameter space to be roughly consistent with previous work by \citet{ggm10}, with the combination of high mass and small semi-major axis most strongly ruled out. However, we find higher upper mass limits than previous studies. This difference almost certainly stems from the increased sampling density of the parameter space. We find that the level of perturbation of the S2 orbit is extremely sensitive to even those parameters traditionally considered to be nuisance parameters, such as the initial mean anomaly of the orbit. 

\begin{figure}[ht]
    \centering
    \resizebox{\hsize}{!}{
    \includegraphics{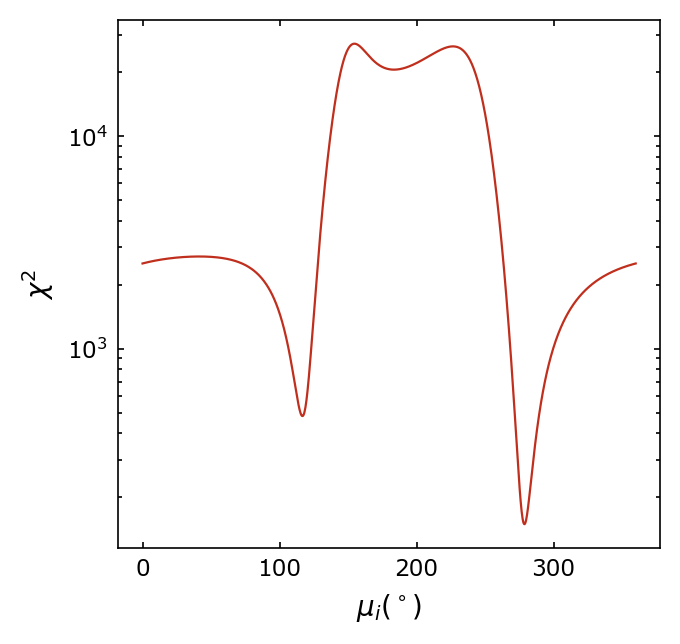}
    }
    \caption{$\chi^2$ vs. the initial mean anomaly of the IMBH. All other IMBH parameters are fixed to the values shown in Table~\ref{tab:imbh_param} and used in Fig.~\ref{fig:valid_imbh}. }
    \label{fig:Lmu}
\end{figure}

We also find a larger allowed region of the parameter space than \citet{nao+20}. We attribute this discrepancy to the fact that the authors in that study approximate the perturbation of the S2 orbit by averaging over the orbital periods of both S2 and the IMBH. As shown in Fig.~\ref{fig:Lmu}, the relative location of the IMBH along its orbit can play a crucial role in determining to what extent it perturbs the path of S2.

\section{Constraints on IMBHs in the GC from the S-stars}
\label{sec:analysis}
In the previous section we report that certain IMBHs with specific orbital properties cannot be excluded given the current GRAVITY and SINFONI/KECK/GNIRS data of S2. In order to understand the long-term effects of such an IMBH, we place it among the 40 S-stars with known orbital parameters \citep[see][]{gil+09, gil+17} and evolve the entire system backwards in time. We essentially run the same simulation as defined in Section~\ref{sec:integration}, but without the posterior sampling. The question we pose is whether the presence of an intermediate-mass perturber destabilises or even disrupts the S-stars within one million years. 

We extract for each of the two scenarios 60 random IMBH orbits from within the 98.8\% likelihood contours shown in Fig.~\ref{fig:constraints_small_large}. Then we evolve the entire system of S2, the 40 S-stars, the IMBH and \sgr with REBOUND/REBOUNDx for $10^6$~years backwards in time. The stars are considered to be `active particles' in the simulation (i.e. they have masses): eight early-type stars have precisely determined masses that lie between 7 and  14~\Msol \citep[see][]{hab+17}; the mass of the lesser-known early-type stars were set to 10~\Msol; the inferred mass of the population of late-type stars lies between 0.5 and 2~\Msol \citep[see][]{hab+19}, and accordingly we set the known late-type stars to 1~\Msol; as the majority of the stellar sample are early-type stars, we also assume   a mass of 10~\Msol for the two stars (S39 and S55) of unidentified spectral type. 

Our criterion to define an unstable system is that within 1~Myr at least one star is ejected and moves past the stellar WR/O disc to reach a separation $r > 3000''$ (about 120~pc) from \sgr. At this distance the stars are far outside the sphere of influence, which has for \sgr a radius of about 3~pc, and appear completely dissociated from the S-star cluster. Depending on the strength of interaction with the IMBH, some of the ejected stars may return to \sgr after increasingly long intervals of time (and on severely modified orbits) which are, however, not covered by our simulation. 

We find that \emph{all} of our IMBH solutions introduce some degree of instability among the S-stars such that their orbits deviate substantially from the non-IMBH case. Furthermore, the majority of our IMBH solutions fulfil our instability criterion: at least one star (but typically several stars) becomes unbound and is ejected well before one million years have passed. The S-stars that strongly interact with an IMBH are in particular the highly eccentric stars such as S9, S14, and S29 with $e > 0.9$. Only a small fraction of about 5\% and 1.6\% of the inner and outer IMBH solutions, respectively, does not disrupt the S-stars in 1~Myr. The stability of the S-star cluster thus gives   a more stringent constraint than the  best-fitting S2 orbit alone.

In our sample of 60 inner IMBH configurations, the only three non-disruptive inner solutions for semi-major axes $0.01'' < a_{\rm i} < 0.1''$ (labelled IMBH$_{\rm i1}$, IMBH$_{\rm i2}$, IMBH$_{\rm i3}$) have similar orbital parameters: masses below 2000~\Msol, moderate to high eccentricities, and a significant inclination towards the S2-plane of at least 60$^{\circ}$. We show their orbital properties together with the only non-disruptive outer solution (IMBH$_{\rm o1}$) in Table~\ref{tab:imbh_param}. Interestingly, the only valid outer solution we find has a mass and semi-major axis that falls into the parameter range proposed by \citet{mer+09} (i.e. at first glance an IMBH that could potentially thermalise the S-stars in a sufficiently short time). \\
\begin{table}[ht]
    \centering
    \begin{tabular}{| l | c | c | c | >{\columncolor[gray]{0.9}} c |}
     \hline
     Parameter                     & IMBH$_{\rm i1}$ & IMBH$_{\rm i2}$ & IMBH$_{\rm i3}$ & IMBH$_{\rm o1}$ \\
     \hline\hline                  
     $M_{\rm i}$ (\Msol)           & 1282   & 1321   &  1130  & 3226  \\
     \hline
     $a_{\rm i}$ ($''$)            & 0.032  & 0.033  & 0.075  & 0.435 \\
     \hline
     $e_{\rm i}$                   & 0.73   &  0.69  & 0.49 & 0.56   \\
     \hline
     $i_{\rm i}$ ($^{\circ}$)      & 52.29  & 63.85  & 75.31 & 274.03 \\
     \hline
     $\Omega_{\rm i}$ ($^{\circ}$) & 155.42 & 161.59 & 291.45  & 95.95 \\
     \hline
     $\omega_{\rm i}$ ($^{\circ}$) & 195.74 & 171.54 & 156.02 & 180.71\\
     \hline
    \end{tabular}
    \vspace{0.25cm}
    \caption{Example solutions of allowed IMBH parameters that do not disrupt the S-star cluster in 1~Myr. The IMBH$_{\rm i1}$ - IMBH$_{\rm i3}$ solutions lie inside the S2 orbit, while IMBH$_{\rm o1}$ is outside.}
    \label{tab:imbh_param}
\end{table}
\begin{figure*}[ht!]
    \includegraphics[width=\textwidth]{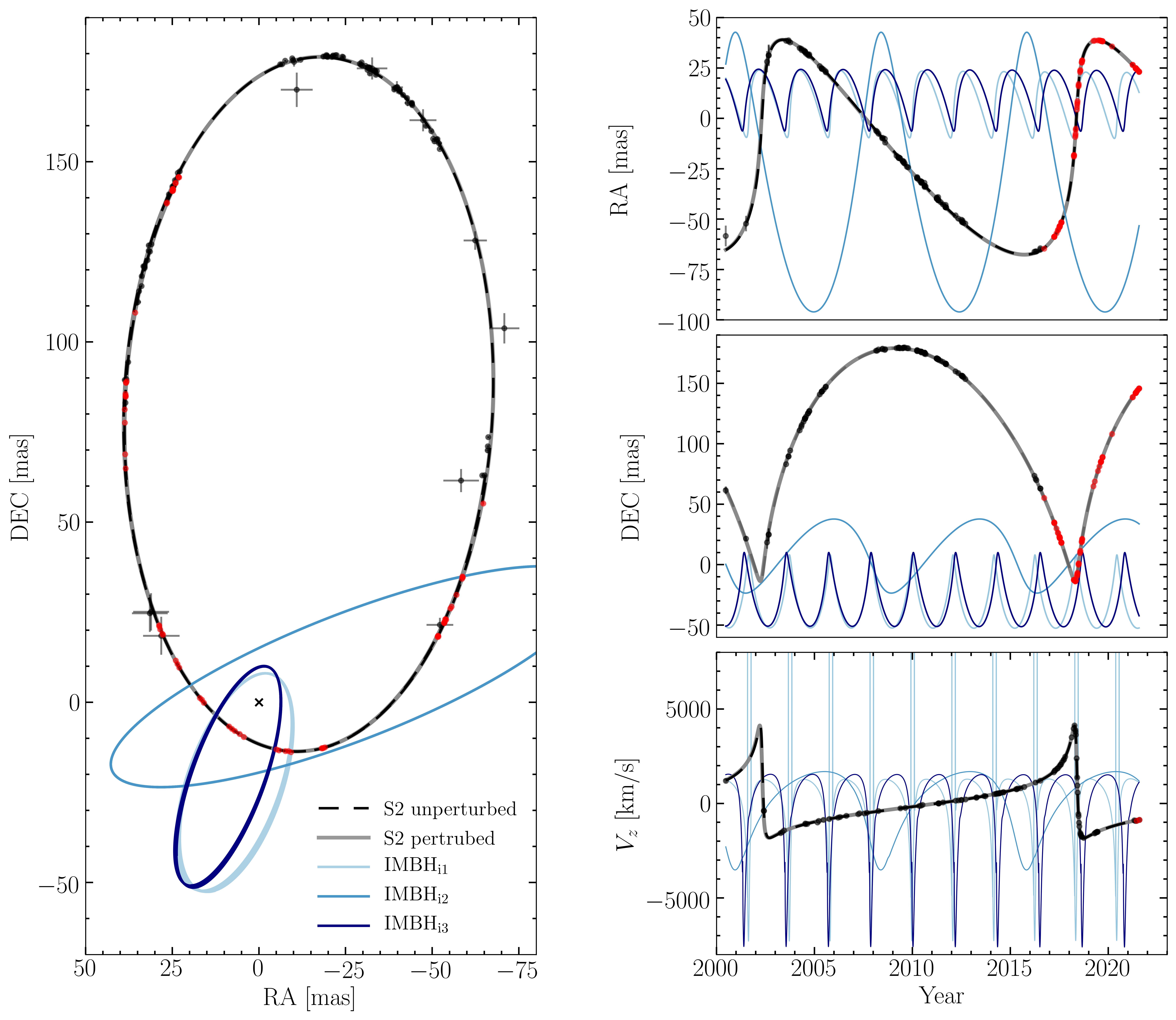}
    \caption{Example orbits of allowed IMBHs in the Galactic Centre. The left panel shows the on-sky orbits of S2 and three IMBH solutions around \sgr (indicated by the cross). The right panels show the time evolution of the RA, DEC, and radial velocity. The solid grey and dashed black curves show the orbit of S2 with and without an IMBH, respectively. The IMBHs are shown in blue and correspond to the parameters given in Table~\ref{tab:imbh_param}. The data points show the last 30 years of observations of S2. The black points correspond to adaptive optics measurements with NACO and early speckle imagery with SHARP. The red points correspond to GRAVITY interferometric measurements. The black and red radial velocity observations correspond, respectively, to SINFONI -- KECK and GNIRS spectral measurements of the line-of-sight velocity.}
    \label{fig:valid_imbh}
\end{figure*}

The three stable inner IMBH orbits are shown in Fig.~\ref{fig:valid_imbh}. We note that we have included the adaptive optics positions measured with the NACO instrument in the plot, although these data points were not used for fitting. The three IMBH orbits shown in blue correspond to the IMBH orbital properties given in Table~\ref{tab:imbh_param} and their residuals are given in Fig.~\ref{fig:residuals}. They demonstrate where and how an IMBH could hide in the GC based on the current GRAVITY and SINFONI/KECK/GNIRS data: the IMBH must have a rather low mass and be on a short orbit around \sgr that is sufficiently inclined towards the orbital plane of S2.

\section{Discussion}
\label{sec:discussion}
Intermediate-mass black holes are thought to play a vital role in the growth of massive and supermassive BHs. They are thus closely linked to the formation and evolution of their host galaxies and are predicted to be abundant in the local universe (e.g. in young dense stellar clusters and dwarf galaxies). However, IMBHs are notoriously difficult to find and unambiguously identify.\footnote{Most IMBHs are thought to roam about alone and rarely accrete matter. Candidate IMBHs are almost exclusively found indirectly as the gravitational source of hyper-velocity compact clouds and hyper-velocity stars, irregular stellar and pulsar dynamics, or in transient ultra-luminous X-ray sources.} The presence or absence of an IMBH in the centre of the Milky Way could give important hints to constrain their formation channel and provide valuable input for future electromagnetic and gravitational wave observations with the Extremely Large Telescope \citep[ELT, e.g.][]{dav+18} and the Laser Interferometer Space Antenna \citep[LISA, planned launch date in 2037; see e.g. ][]{LISA17}, respectively.\\

In this paper we used the high angular resolution astrometric and spectroscopic data of the star S2 from GRAVITY and SINFONI/KECK/GNIRS, respectively, to assess where in the GC an IMBH could hide. We had a fresh look at the dynamical search for IMBHs in the GC by exploring the full 16-dimensional parameter space of the chaotic three-body problem comprising \sgr, an IMBH, and the star S2. We specifically considered two scenarios, one where the IMBH trajectory lies inside the S2 orbit with a semi-major axis $0.01'' < a_{\rm i} < 0.1''$ and the other where the IMBH trajectory crosses the S2 orbit or lies outside, $0.1'' < a_{\rm i} < 1''$, and calculated for both cases the resulting modified orbital properties for S2. Using dynamic nested sampling, we explored the full set of parameters and found for each scenario the most likely locations for an IMBH (see Fig.~\ref{fig:constraints_small_large}). 

We found that for very specific combinations of orbital parameters, in particular for certain IMBH orientations and pericentre passage times, high IMBH masses could be located among the S-stars. This happens for IMBHs that stay sufficiently far from S2 during their closest approach to \sgr so as not to measurably affect the orbit at all. We therefore analysed our valid solutions further and selected for each scenario 60 random solutions from within the 98.8\% likelihood contours (see Fig.~\ref{fig:constraints_small_large}). These IMBHs were placed among the 40 stars of the S-star cluster and evolved backwards in time for one million years. Moreover, we calculated for each set of 60 solutions the residuals between the data and the models. 

Based on the results from the optimisation, the stability analysis and the residual calculation, we arrive at the conclusion that although we find viable fits to the data that suggest an IMBH could be present for specific parameter combinations, the majority of these solutions do not withstand the reality check and would disrupt the S-star cluster in less than a million years or induce a precession in the orbit of S2 beyond the observed one. We conclude the following:

\begin{itemize}    
    \item Current GRAVITY and SINFONI/KECK/GNIRS data do not formally require the presence of an IMBH.
    \item IMBHs on orbits that cross the S2 orbit or lie outside the S2 orbit among the other S-stars are disfavoured as they typically disrupt the S-star cluster in less than one million years (only 1.6\% of the solutions with $0.1'' < a_{\rm i} < 1''$ are stable).
    \item A low-mass IMBH with M < 2000~\Msol could hide inside the S2 orbit if its orbit is sufficiently inclined towards S2 (only 5\% of the solutions with $0.01'' < a_{\rm i} < 0.1''$ are stable, all of them low-mass IMBHs).
\end{itemize}

We conclude the following from the IMBH constraints on the population(s) of stars and stellar remnants in the GC: A spherical distribution of stellar-mass BHs, neutron stars (NSs) and/or white dwarfs (WDs) located as a dark cluster among the S-stars would be affected by an IMBH in a very similar way to the S-stars. The bodies on eccentric orbits would most likely be ejected, leaving preferentially the compact objects on low-eccentricity orbits behind. The total mass of such a dark (extended) cluster has been constrained to about 15000~\Msol within the S-star cluster. Conversely, in the absence of an IMBH among the S-stars, which is based on our analysis the preferred case, a dark cluster of SBHs, NSs, and/or WDs could show a wide range of eccentricities and orbital inclinations, and thus exhibit morphological similarities to the S-star cluster. \\


The high-precision GRAVITY astrometric measurements span at present about half of the S2 orbit. Much stronger constraints on the properties of IMBHs can be obtained once GRAVITY has measured a full S2 orbit. Already the current SINFONI/KECK/GNIRS data which cover, albeit sparsely, the 2002 pericentre passage hint that two consecutive pericentre passages will be invaluable to assess the likelihood of a low-mass IMBH on a $ < 0.1''$ orbit. After the  upcoming closest approach of S2 to \sgr in 2034, the data will allow us to put stronger constraints on a single IMBH companion of \sgr as well as its extended mass \citep[see][]{GRAVITYcol22, hei+22, rub+01}. Moreover, there are now several other stars with complete or near-complete orbits that can already serve   in the coming few years as additional precision probes. Knowing whether or not the nuclear cluster in the GC hosts an IMBH will in turn put constraints on the formation processes of IMBHs. Moreover, constraints on the mass distribution in the GC will also be of value to LISA, which will be able to measure gravitational waves of moving masses in the GC.


\begin{acknowledgements}
We are very grateful to our funding agencies (MPG, ERC, CNRS [PNCG, PNGRAM], DFG, BMBF, Paris Observatory [CS, PhyFOG], Observatoire des Sciences de l’Univers de Grenoble, and the Fundação para a Ciência e Tecnologia) and to ESO. We especially thank the excellent and in every way amazing ESO/Paranal staff as well as the scientific and technical staff members in our institutions who helped to make GRAVITY and SINFONI a reality and observations a success. P.G. and V.C. were supported by Fundação para a Ciência e a Tecnologia, with grants reference SFRH/BSAB/142940/2018, UIDB/00099/2020 and PTDC/FIS-AST/7002/2020. S.G. acknowledges the support from ERC starting grant No. 306311. F.E. acknowledges the support from ERC synergy grant No. 610058. The GNIRS spectra were obtained at the international Gemini Observatory, a program of NSF’s NOIRLab, managed by the Association of Universities for Research in Astronomy (AURA) under a cooperative agreement with the National Science Foundation (NSF) on behalf of the Gemini Observatory partnership: the National Science Foundation (United States), National Research Council (Canada), Agencia Nacional de Investigación y Desarrollo (Chile), Ministerio de Ciencia, Tecnología e Innovación (Argentina), Ministério da Ciência, Tecnologia, Inovações e Comunicações (Brazil), and Korea Astronomy and Space Science Institute (Republic of Korea). This work was enabled by observations made from the Gemini North telescope, located within the Maunakea Science Reserve and adjacent to the summit of Maunakea. We are grateful for the privilege of observing the Universe from a place that is unique in both its astronomical quality and its cultural significance.
\end{acknowledgements}

\bibliographystyle{aa}
\bibliography{GC_IMBH.bib}

\begin{appendix}
\section{The full IMBH parameter space}
\label{app:fullparam}
\begin{figure}[hb!]
   \includegraphics[width=\textwidth]{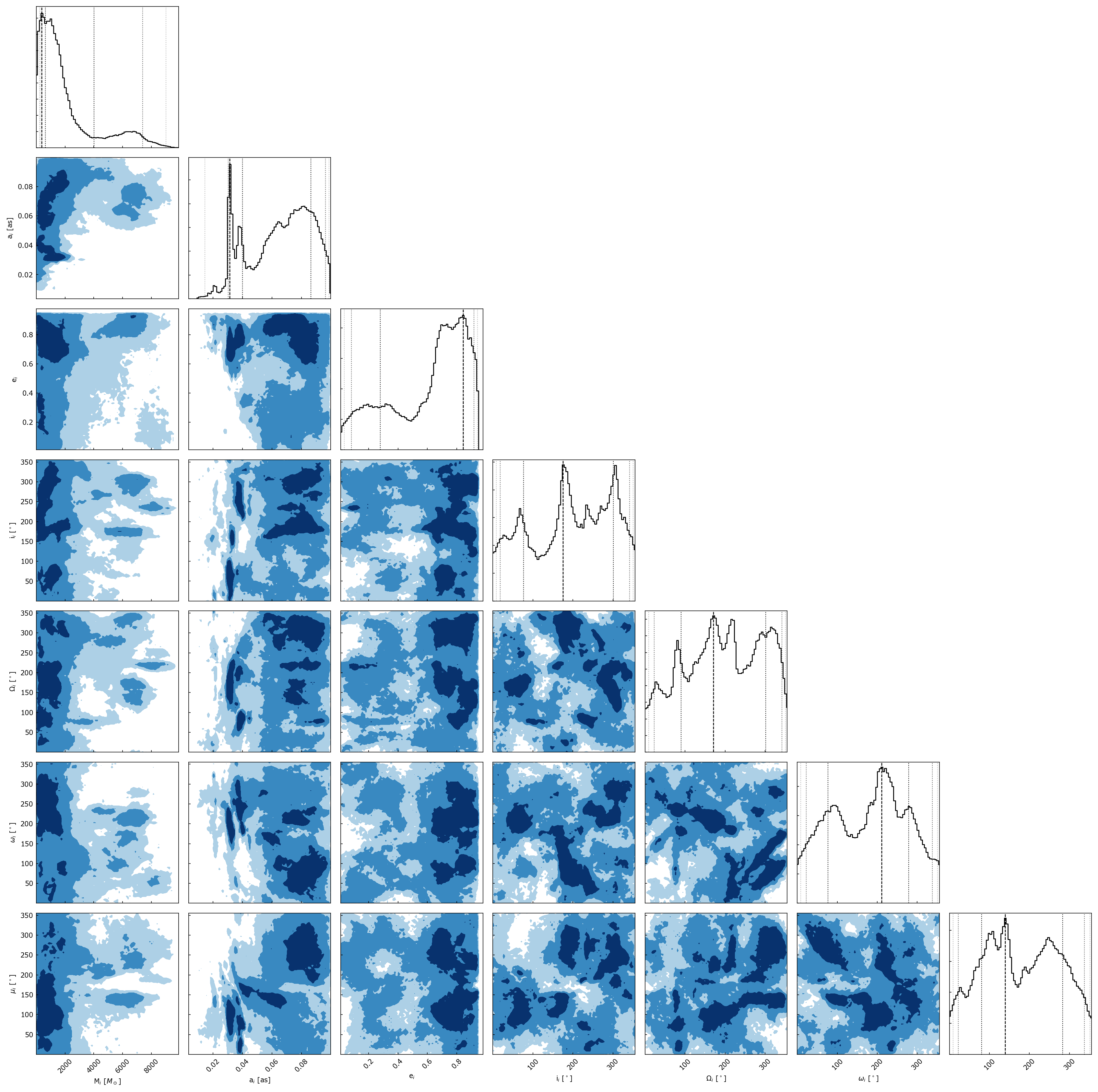}
    \caption{Corner plot of the posterior of the IMBH orbital parameters for the hierarchical case. Not shown are the parameters of the S2 orbit, which are also allowed to vary. }
    \label{fig:IMBH_Corner}
\end{figure}

Figure~\ref{fig:IMBH_Corner} shows the posterior over all of the IMBH orbital parameters. For clarity in plotting, we do not show the parameters of the S2 orbit, which are also allowed to vary over the fit. We find that the S2 parameters are tightly constrained and that their values match those found in previous works \citep[e.g.][]{GRAVITYcol18a}. 

\clearpage

\section{Residuals}
\begin{figure}[hb!]
    \includegraphics[width=\columnwidth]{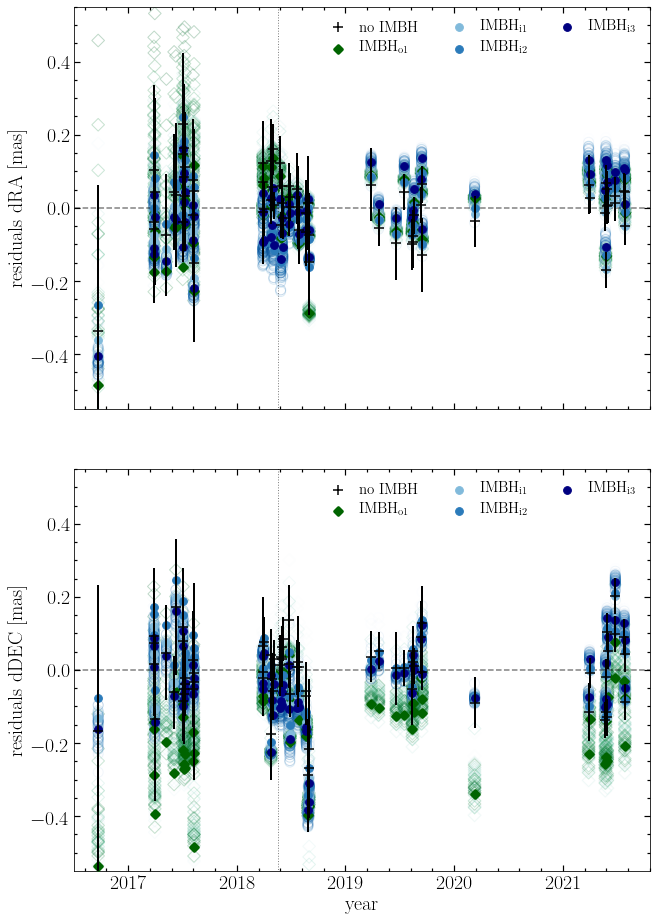}
    \caption{Residuals of the S2 orbit. The best-fitting Schwarzschild model (i.e. without the presence of an IMBH) is marked in black. The blue open circles denote the 60 solutions where the IMBH orbit lies inside the S2 orbit, $0.01'' < a_{\rm i} < 0.1''$, and the green open diamonds are   the 60 outer IMBH solutions with $0.1'' < a_{\rm i} < 1''$. The filled symbols highlight solutions shown in Table~\ref{tab:imbh_param}. The fine vertical line indicates the time of pericentre passage. Upper panel: Data--model difference for the right ascension (dRA). Bottom panel: Data--model difference for the declination (dDEC).}
    \label{fig:residuals}
\end{figure}

We calculate the residuals between the S2 data and the model orbits for all 60 inner and outer IMBH solutions. Figure~\ref{fig:residuals} compares the astrometry residuals of the unperturbed Schwarzschild orbit (black markers) of S2 to the three orbits that are modified by an inner IMBH (blue dots) and the orbit modified by an outer IMBH (green diamonds). The entirety of all residuals, including orbits that have been rejected based on stability arguments, are shown as a open symbols in the background. Both the inner and the outer stable IMBH solutions have residuals comparable to a pure Schwarzschild orbit. 

\end{appendix}

\end{document}